\documentclass[a4paper]{jpconf}
\usepackage{graphicx}
\def\bb#1{\mbox{\footnotesize $(#1)$}}
\def\eqref#1{(\ref{#1})}

\begin{document}
\title{Phase-space continuity equations for quantum decoherence, purity, von Neumann and R\'enyi entropies}

\author{Alex E. Bernardini}
\address{Departamento de F\'{\i}sica, Universidade Federal de S\~ao Carlos, PO Box 676, 13565-905, S\~ao Carlos, SP, Brasil.}
\ead{alexeb@ufscar.br}
\author{Orfeu Bertolami}
\address{Departamento de F\'isica e Astronomia, Faculdade de Ci\^encias da Universidade do Porto, Rua do Campo Alegre, 687, 4169-007, Porto, Portugal}
\ead{orfeu.bertolami@fc.up.pt}

\begin{abstract}
Phase-space features of the Wigner flow are examined so to provide a set of continuity equations that describe the flux of quantum information in the phase-space.
The reported results suggest that the non-classicality profile of anharmonic (periodic) quantum systems can be consistently obtained in terms of the fluxes of probability, purity and von Neumann entropy.
Extensions of the such phase-space quantifiers are also investigated in the context of the so-called R\'enyi entropy.
\end{abstract}

\section{Introduction}

\hspace{1 em} The formulation of the phase-space dynamics of quantum systems according to the Weyl-Wigner quantum mechanics \cite{Wigner,Case} provides elementary tools to connect the predictive features of quantum mechanics with the investigation of quantum information.
As it has been shown, the Weyl-Wigner phase-space formulation \cite{01A,02A,03A} can provide a wider understanding of the limits between classical and quantum regimes of physical systems.

The Wigner function, $W(q, p)$, encompasses the information from coordinate and momentum Schr\"odinger wave function representation of quantum mechanics in a convolutional representation of the wave function described by the Weyl transform of a quantum operator, $\hat{Q}$,
\begin{equation}
Q^W(q, p)
= \hspace{-.2cm} \int^{+\infty}_{-\infty} \hspace{-.15cm}ds\,\exp{\left[2\,i \,p\, s/\hbar\right]}\,\langle q - s | \hat{Q} | q + s \rangle=\hspace{-.2cm} \int^{+\infty}_{-\infty} \hspace{-.15cm} dr \,\exp{\left[-2\, i \,q\, r/\hbar\right]}\,\langle p - r | \hat{Q} | p + r\rangle,
\end{equation}
such that 
\begin{equation}
 h^{-1} \hat{\rho} \to W(q, p) = (\pi\hbar)^{-1} 
\int^{+\infty}_{-\infty} \hspace{-.15cm}ds\,\exp{\left[2\, i \, p \,s/\hbar\right]}\,
\psi(q - s)\,\psi^{\ast}(q + s),
\end{equation}
the Weyl transform of a density matrix, $\hat{\rho} = |\psi \rangle \langle \psi |$, is interpreted as the Fourier transform of the off-diagonal elements of $\hat{\rho}$.

In spite of being consistent with the Moyal's picture of quantum mechanics \cite{Moyal}, where the noncommutative nature of canonical operators supports the Moyal {\em star}-product definition, the above definition does not yield a real-valued probability distribution, given that it admits local negative values, which is inconsistent with the positive-valued probability interpretation. This provides a {\em quasi}-probability interpretation which admits an enhanced phase-space framework and eventually fixes this issue \cite{Ballentine}.

Keeping in mind that the Weyl transform and the Wigner function connect quantum observables, $\hat{Q}$, with their expectation values, one notices that the trace of the product of the operators, $\hat{\rho}$ and $\hat{Q}$, can be evaluated from the $dq\,dp$ integral of the product of their Weyl transforms \cite{Wigner,Case},
\footnote{In general one has \begin{equation}Tr_{\{q,p\}}\left[\hat{Q}_1\hat{Q}_2\right] = \int^{+\infty}_{-\infty} \hspace{-.15cm}\int^{+\infty}_{-\infty} \hspace{-.15cm} {dq\,dp} \,{Q}^W_1(q, p)\,{Q}^W_2(q, p),\end{equation} for any product of two operators, $\hat{Q}_{1}$ and $\hat{Q}_{2}$.}
\begin{equation}
Tr_{\{q,p\}}\left[\hat{\rho}\hat{Q}\right] \to \langle Q \rangle = 
\int^{+\infty}_{-\infty} \hspace{-.15cm}\int^{+\infty}_{-\infty} \hspace{-.15cm} {dq\,dp}\,W(q, p)\,{{Q}^W}(q, p),
\label{zzvfive}
\end{equation}
which allows for the probability distribution interpretation, with the normalization condition established by $\hat{\rho}$, $Tr_{\{q,p\}}[\hat{\rho}]=1$.
The Weyl-Wigner definitions also admit extensions from pure states to statistical mixtures. In particular, the purity,
\begin{equation}
Tr_{\{q,p\}}[\hat{\rho}^2] = 2\pi\int^{+\infty}_{-\infty} \hspace{-.15cm}\int^{+\infty}_{-\infty} \hspace{-.15cm} {dq\,dp}\,W(q, p)^2,
\label{zzvpureza}
\end{equation}
is constrained by the conditions of $Tr_{\{q,p\}}[\hat{\rho}^2] = Tr_{\{q,p\}}[\hat{\rho}] = 1$ for pure states. In the investigation of classical and quantum regimes of one-dimensional physical systems, the statistical aspects related to the nature of the density matrix quantum operators play a relevant role in the computation of entropy and information quantifiers. 

Considering these theoretical foundations, a fluid-analog of the phase-space quantum entropy (information) flux in the phase-space can be recovered from the Weyl-Wigner formalism of quantum mechanics \cite{EPL18}.
In this contribution, phase-space continuity equations for quantum decoherence, purity and von Neumann entropy are recast in a dimensionless framework in order to be useful for discussing a broad set of Hamiltonian quantum systems.
In particular, extensions of such phase-space quantifiers are also investigated in the context of the R\'enyi entropy definition.
The conclusion is that, the set of phase-space continuity equations discussed below emphasizes the relevancy of the Wigner formalism in discussing the flux of quantum information in the phase-space.

\section{Phase-space flow analysis and continuity equations}

\hspace{1 em} The dynamics of the Wigner function, $W(q,\,p;\,t)$ can be described in terms of a vector flux, $\mathbf{J}(q,\,p;\,t)$, which represents the flow of $W(q,\,p;\,t)$ in the phase-space \cite{02A,Ferraro11,Donoso12,04}.
In this case, the flow field, $\mathbf{J}(q,\,p;\,t)$, is decomposed into coordinate and momentum components, $\mathbf{J} = J_q\,\hat{q} + J_q\,\hat{q}$, with $\hat{p} = \hat{p}_q$, through which one can identify the quantum version of the Liouville equation, that is a continuity equation given by \cite{Case,Ballentine,02A,EPL18},
\begin{equation}
\frac{\partial W}{\partial t} + \frac{\partial J_q}{\partial q}+\frac{\partial J_p}{\partial p} \equiv
\frac{\partial W}{\partial t} + \mbox{\boldmath $\nabla$}\cdot \mathbf{J} =0,
\label{zzvquaz51}
\end{equation}
where \begin{equation}
J_q(q,\,p;\,t)= \frac{p}{m}\,W(q,\,p;\,t), \label{zzvquaz500BB}
\end{equation}
\begin{equation}
J_p(q,\,p;\,t) = -\sum_{\nu=0}^{\infty} \left(\frac{i\,\hbar}{2}\right)^{2\nu}\frac{1}{(2\nu+1)!} \, \left[\left(\frac{\partial~}{\partial q}\right)^{2\nu+1}\hspace{-.5cm}V(q)\right]\,\left(\frac{\partial~}{\partial p}\right)^{2\nu}\hspace{-.3cm}W(q,\,p;\,t),
\label{zzvquaz500}
\end{equation}
and $V(q)$ is the potential for a non-relativistic quantum system described by the Hamiltonian
\begin{equation}
H(q,\,p) = \frac{p^2}{2m} + V(q),
\end{equation}
Notice from Eq.~\eqref{zzvquaz500} that the contributions from $j \geq 1$ in the series expansion quantify the distortion due to the quantum corrections with respect to the classical Liouvillian phase-space trajectories.

In particular, for the generalized discussion of non-relativistic quantum systems
where $m$ is simply a mass scale, generalized results can be obtained when $H(q,\,p)$ is put into a dimensionless form, $\mathcal{H}(x,\,k) = k^2/2 + \mathcal{U}(x)$.
The dimensionless variables, $x = \left(m\,\omega\,\hbar^{-1}\right)^{1/2} q$ and $k = \left(m\,\omega\,\hbar\right)^{-1/2}p$, provide the identification of $\mathcal{H} = (\hbar \omega)^{-1} H$ and $\mathcal{U}(x) = (\hbar \omega)^{-1} V\left(\left(m\,\omega\,\hbar^{-1}\right)^{-1/2}x\right)$, where $\omega^{-1}$ is the time scale.

The corresponding dimensionless form of the Wigner function and of the corresponding Wigner current are thus obtained from
\begin{eqnarray}
\mathcal{W}(x, \, k;\,\tau) &\equiv& \left(m\omega\hbar\right)^{1/2}\, W(q,\,p;\,t),\\
\mathcal{J}_x(x, \, k;\,\tau) &\equiv& m \,\, J_q(q,\,p;\,t),\\
\mathcal{J}_k(x, \, k;\,\tau) &\equiv& \omega^{-1}\, J_p(q,\,p;\,t),
\end{eqnarray}
which can be recast in the form \cite{NovoPaper}
\small\begin{eqnarray}\label{zzvDimW}
\mathcal{W}(x, \, k;\,\tau) &=& \pi^{-1} \int^{+\infty}_{-\infty} \hspace{-.15cm}dy\,\exp{\left[2\, i \, k \,y\right]}\,\varphi(x - y;\,\tau)\,\varphi^{\ast}(x + y;\,\tau),\quad \mbox{with $y = \left(m\,\omega\,\hbar^{-1}\right)^{1/2} s$},\,\,\,\,\\
\label{zzvDimWA}\mathcal{J}_x(x, \, k;\,\tau) &=& k\,\mathcal{W}(x, \, k;\,\tau)
,\\
\label{zzvDimWB}\mathcal{J}_k(x, \, k;\,\tau) &=& -\sum_{\nu=0}^{\infty} \left(\frac{i}{2}\right)^{2\nu}\frac{1}{(2\nu+1)!} \, \left[\left(\frac{\partial~}{\partial x}\right)^{2\nu+1}\hspace{-.5cm}\mathcal{U}(x)\right]\,\left(\frac{\partial~}{\partial k}\right)^{2\nu}\hspace{-.3cm}\mathcal{W}(x, \, k;\,\tau),
\end{eqnarray}\normalsize
where $\tau = \omega t$ is a dimensionless quantity\footnote{Notice that $\varphi(x,\,\tau)$ is consistent with the normalization condition given by
\begin{equation}
\int^{+\infty}_{-\infty} \hspace{-.2 cm}{dx}\,\vert\varphi(x;\,\tau)\vert^2 =\int^{+\infty}_{-\infty}\hspace{-.2 cm}{dq}\,\vert\psi(q;\,t)\vert^2 = 1,
\end{equation}
and the Eq.~\eqref{zzvquaz51} can be multiplied by $(m\hbar/\omega)^{1/2}$ so to yield the dimensionless continuity equation,
\begin{equation}
\frac{\partial \mathcal{W}}{\partial \tau} + \frac{\partial \mathcal{J}_x}{\partial x}+\frac{\partial \mathcal{J}_k}{\partial k} = \frac{\partial \mathcal{W}}{\partial \tau} + \mbox{\boldmath $\nabla$}_{\xi}\cdot\mbox{\boldmath $\mathcal{J}$} =0,
\end{equation}
where the phase-space coordinate vector, $\mbox{\boldmath $\xi$} = (x,\,k)$, is defined.}.
From this framework, the phase-space information flux associated to extensions of the above continuity equation can be obtained \cite{JCAP18,EPL18}.

To implement a self-contained procedure for obtaining the continuity equations derived from the fluid-analog properties of the Wigner functions, one notices that locally and globally conservative properties associated to a point in the phase-space, $\mbox{\boldmath $\xi$}$, and to a phase-space {\em volume integral bounded by a comoving closed surface}, $V = \int_{V}dx\,dk$, are connected by a convective (or substantial) derivative \cite{02A,EPL18} operator defined by
\begin{equation}
\frac{D~}{D\tau} \int_{V}dV\,\mathcal{W} \equiv 
\int_{V}dV\,\left[\frac{D\mathcal{W}}{D\tau} + \mathcal{W} \mbox{\boldmath $\nabla$}_{\xi}\cdot \mathbf{v}_{\xi}\right]\label{zzvquaz57D},
\end{equation}
with $dV \equiv dx\,dk$, and where $ \mathbf{v}_{\xi}$ corresponds to the classical phase-space vector velocity, $\mathbf{v}_{\xi(\mathcal{C})} = (k,\, -\partial \mathcal{U}/\partial x)$.
Assuming that a two-dimensional classical path, $\mathcal{C}$, can be identified, from the definition of $\mathbf{v}_{\xi(\mathcal{C})}$, one sets
\begin{equation}
\frac{D \mathcal{W}}{D\tau} = - \mathcal{W}\, \mbox{\boldmath $\nabla$}_{\xi} \cdot \mathbf{v}_{\xi(\mathcal{C})},
\label{zzvquaz57B}
\end{equation}
which implies into a conservation law, ${D \mathcal{W}}/{D\tau} = 0$.
In this case, the divergenceless behavior of the classical fluid-analog associated to the Wigner function is expressed by $\mbox{\boldmath $\nabla$}_{\xi} \cdot \mathbf{v}_{\xi(\mathcal{C})} = 0$, which defines the Liouvillian behavior.

To implement the above properties onto a quantum scenario, one has to notice that $\mbox{\boldmath$\mathcal{J}$}$ can be cast in the form of product $\mathbf{w}\,\mathcal{W}$, where the now introduced Wigner phase-velocity, $\mathbf{w}$, satisfies the constraint equation given by
\begin{equation}
\mbox{\boldmath $\nabla$}_{\xi} \cdot \mathbf{w} = \frac{\mathcal{W}\, \mbox{\boldmath $\nabla$}_{\xi}\cdot \mbox{\boldmath$\mathcal{J}$} - \mbox{\boldmath$\mathcal{J}$}\cdot\mbox{\boldmath $\nabla$}_{\xi}\mathcal{W}}{\mathcal{W}^2} \neq0,
\label{zzvquaz59}
\end{equation}
that can be used to quantify the departure from the Liouvillian behavior \cite{02A,EPL18}.

For Hamiltonians describing periodic motions (in particular, for those driven by symmetric quantum wells), the parameterization to the two-dimensional volume enclosed by the classical path, $\mathcal{C}$, results into
\begin{equation}
\sigma_{(\mathcal{C})} =\int_{V_{_{\mathcal{C}}}}dV\,\mathcal{W},
\label{zzvquaz60}
\end{equation}
where $\sigma_{(\mathcal{C})}$ corresponds to the phase-space volume integrated probability flux.

Using the properties of $\mathbf{w}$, after some manipulation involving Eqs.~(\ref{zzvquaz57D})-(\ref{zzvquaz57B}), one obtains
\begin{equation}
 \frac{D~}{D\tau}\sigma_{(\mathcal{C})} =\frac{D~}{D\tau}\int_{V_{_{\mathcal{C}}}}dV \,\mathcal{W} = \int_{V_{_{\mathcal{C}}}}dV \,\left[\mbox{\boldmath $\nabla$}_{\xi}\cdot (\mathbf{v}_{\xi(\mathcal{C})}\mathcal{W}) - \mbox{\boldmath $\nabla$}_{\xi}\cdot \mbox{\boldmath$\mathcal{J}$}\right],
\label{zzvquaz51CC}
\end{equation}
which allows for identifying the role of the quantum corrections
given in terms of $\Delta \mbox{\boldmath$\mathcal{J}$} = \mbox{\boldmath$\mathcal{J}$} - \mathbf{v}_{\xi(\mathcal{C})}\mathcal{W}$, which effectively drives the volume variation of $\sigma_{(\mathcal{C})}$ in terms of a path integral given by
\begin{equation}
 \frac{D~}{D\tau}\sigma_{(\mathcal{C})} = -\int_{V_{_{\mathcal{C}}}}dV\, \mbox{\boldmath $\nabla$}_{\xi}\cdot \Delta \mbox{\boldmath$\mathcal{J}$} = -\oint_{\mathcal{C}}d\ell\, \Delta\mbox{\boldmath$\mathcal{J}$}\cdot \mathbf{n}\equiv -\oint_{\mathcal{C}}d\ell\, \mbox{\boldmath$\mathcal{J}$}\cdot \mathbf{n},
\label{zzvquaz51DD}
\end{equation}
where the unitary vector $\mathbf{n}$ is defined by $\mathbf{n}= (-d{k}_{_{\mathcal{C}}}/d\tau, d{x}_{_{\mathcal{C}}}/d\tau) \vert\mathbf{v}_{\xi(\mathcal{C})}\vert^{-1}$ and $\mathbf{n}\cdot\mathbf{v}_{\xi(\mathcal{C})}= 0$ has been used in the last step. One thus has a parametric integral given by
\begin{equation}
\frac{D~}{D\tau}\sigma_{(\mathcal{C})}
\bigg{\vert}_{\tau = T} = -\oint_{\mathcal{C}}d\ell\, \Delta\mbox{\boldmath$\mathcal{J}$}\cdot \mathbf{n} = -
\int_{0}^{T}d\tau\, \Delta \mathcal{J}_p(x_{_{\mathcal{C}}}\bb{\tau},\,k_{_{\mathcal{C}}}\bb{\tau};\tau)\,\,\frac{d}{d\tau}{x}_{_{\mathcal{C}}}\bb{\tau},
\label{zzvquaz51EE}
\end{equation}
where, in the first step, the line element, $\ell$, is set as $d\ell \equiv \vert\mathbf{v}_{\xi(\mathcal{C})}\vert d\tau$, and, in the second step, one has $x_{_{\mathcal{C}}}\bb{\tau}$ and $k_{_{\mathcal{C}}}\bb{\tau}$ as typical solutions of the classical Hamiltonian problem. In addition, $T=2\pi$ is the dimensionless period of the classical motion, and $\Delta \mathcal{J}_p(x,\,p;\tau)$ is identified by the series expansion from Eq.~\eqref{zzvquaz500} for $\nu\geq 1$.

The above result straightforwardly quantifies the loss of information related to quantum decoherence.
The same analysis can be applied to the informational content related to the Wigner associated von Neumann entropy and purity expressed respectively by \cite{EPL18}
\begin{equation}
{S}_{vN} =-\int_{V}dV\,\mathcal{W}\,\ln\vert \mathcal{W}\vert,
\label{zzvquaz60}
\end{equation}
and
\begin{eqnarray}
\mathcal{P} = 2\pi \int_{V}dV\,\, \mathcal{W}^2.
\label{zzvquaz63}
\end{eqnarray}

In these cases, some suitable mathematical manipulations involving Eq.~\eqref{zzvquaz59}\footnote{In particular, if one notices that $\mbox{\boldmath $\nabla$}_{\xi}\cdot\mbox{\boldmath$\mathcal{J}$} = \mathcal{W}\,\mbox{\boldmath $\nabla$}_{\xi}\cdot\mathbf{w}+ \mathbf{w}\cdot \mbox{\boldmath $\nabla$}_{\xi}\mathcal{W}$.} result into
\begin{eqnarray}
\frac{D{S}_{vN}}{D\tau} &=& -\frac{D~}{D\tau}\left(\int_{V}dV\, \mathcal{W}\,\ln(\mathcal{W})\right)
\nonumber\\
&=& -\int_{V}dV\,\left[\frac{D~}{D\tau} (\mathcal{W}\,\ln(\mathcal{W})) + \mathcal{W}\,\ln(\mathcal{W}) \mbox{\boldmath $\nabla$}_{\xi}\cdot \mathbf{v}_{\xi(\mathcal{C})}\right]\nonumber\\
&=&- \int_{V}dV\,\left[\frac{\partial~}{\partial \tau} (\mathcal{W}\,\ln(\mathcal{W})) + \mbox{\boldmath $\nabla$}_{\xi}\cdot(\mathbf{v}_{\xi(\mathcal{C})} \mathcal{W}\,\ln(\mathcal{W}))\right]\nonumber\\
&=& \int_{V}dV\,\left[\mathcal{W} \,\mbox{\boldmath $\nabla$}_{\xi}\cdot\mathbf{w} +
 \mbox{\boldmath $\nabla$}_{\xi}\cdot\left(\mbox{\boldmath$\mathcal{J}$} \ln(\mathcal{W}) - \mathbf{v}_{\xi(\mathcal{C})} \mathcal{W}\,\ln(\mathcal{W})\right)\right]\nonumber\\
 &=& \int_{V}dV\,\mathcal{W} \,\mbox{\boldmath $\nabla$}_{\xi}\cdot\mathbf{w}
+ \oint_{\mathcal{\,\,}}d\ell\, \ln(\mathcal{W})\left(\mbox{\boldmath$\Delta\mathcal{J}$}\cdot \mathbf{n}\right)\nonumber\\
&\stackrel{V\to\infty}{=}& \langle \mbox{\boldmath $\nabla$}_{\xi}\cdot\mathbf{w}\rangle,
\label{zzvquaz61}
\end{eqnarray}
and
\begin{eqnarray}
\frac{1}{2\pi}\frac{D\mathcal{P}}{D\tau} &=& \frac{D~}{D\tau}\left(\int_{V}dV\, \mathcal{W}^2\right)\nonumber\\
&=&
\int_{V}dV\,\left[\frac{D~}{D\tau} \mathcal{W}^2 + \mathcal{W}^2 \mbox{\boldmath $\nabla$}_{\xi}\cdot \mathbf{v}_{\xi(\mathcal{C})}\right]\nonumber\\
&=& \int_{V}dV\,\left[\frac{\partial~}{\partial \tau}\mathcal{W}^2 + \mbox{\boldmath $\nabla$}_{\xi}\cdot(\mathbf{v}_{\xi(\mathcal{C})} \mathcal{W}^2)\right]\nonumber\\
&=& -\int_{V}dV\,\left[\mathcal{W}^2 \,\mbox{\boldmath $\nabla$}_{\xi}\cdot\mathbf{w} +
 \mbox{\boldmath $\nabla$}_{\xi}\cdot(\mbox{\boldmath$\mathcal{J}$} \mathcal{W} - \mathbf{v}_{\xi(\mathcal{C})} \mathcal{W}^2)\right]\nonumber\\
&=& \int_{V}dV\,\mathcal{W}^2 \,\mbox{\boldmath $\nabla$}_{\xi}\cdot\mathbf{w}
+ \oint_{\mathcal{\,\,}}d\ell\, \mathcal{W} \left(\mbox{\boldmath$\Delta\mathcal{J}$}\cdot \mathbf{n}\right)\nonumber\\
&\stackrel{V\to\infty}{=}& \langle \mathcal{W} \mbox{\boldmath $\nabla$}_{\xi}\cdot\mathbf{w}\rangle,
\label{zzvquaz64}
\end{eqnarray}
for von Neumann entropy and purity, respectively, where $\langle \dots \rangle = Tr_{\{x,k\}}\left[\hat{\rho}(\dots)\right]$ and, in both cases, it has been used that $\partial \mathcal{W}/\partial \tau = - \mbox{\boldmath $\nabla$}_{\xi}\cdot\mbox{\boldmath$\mathcal{J}$} = - \mbox{\boldmath $\nabla$}_{\xi}\cdot (\mathbf{w}\,\mathcal{W}$).

Notice that the surface terms have been consistently suppressed in the limit where $V \to \infty$, which leads to the results from Refs. \cite{EPL18,NovoPaper}.
However, when a classical surface, ${\mathcal{C}}$, encloses a finite phase-space volume, $V_{_{\mathcal{C}}}$, the surface term must be taken into account.
In fact, given the definitions from Eqs.~\eqref{zzvDimW}-\eqref{zzvDimWB}, it is possible to verify that $\mbox{\boldmath $\nabla$}_{\xi}\cdot\mathbf{w}$ is proportional to $$ \sum_{\nu=1}^{\infty} \left(\frac{i}{2}\right)^{2\nu}\frac{1}{(2\nu+1)!} \, \left[\left(\frac{\partial~}{\partial x}\right)^{2\nu+1}\hspace{-.5cm}\mathcal{U}\right]\,\mathcal{W}^2\frac{\partial~}{\partial k}\left((1/\mathcal{W})\frac{\partial~}{\partial k}\right)^{2\nu}\hspace{-.3cm}\mathcal{W},$$ and for parity symmetric potentials, $\mathcal{U}(x) =\mathcal{U} (-x)$ which lead to periodic anharmonic motions, the above obtained averaged terms vanish. Therefore, the obtained continuity equations can be recast in the form of \cite{EPL18,NovoPaper}
\begin{eqnarray}
\frac{D~}{D\tau}{S}_{vN(\mathcal{C})}
\bigg{\vert}_{\tau = T} &=& \oint_{\mathcal{C}}d\ell\, \ln(\mathcal{W})\left(\Delta\mbox{\boldmath$\mathcal{J}$}\cdot \mathbf{n}\right) \nonumber\\
&=&
\int_{0}^{T}d\tau\, \ln(\mathcal{W}(x_{_{\mathcal{C}}}\bb{\tau},\,k_{_{\mathcal{C}}}\bb{\tau};\tau))\,\,\Delta \mathcal{J}_p(x_{_{\mathcal{C}}}\bb{\tau},\,k_{_{\mathcal{C}}}\bb{\tau};\tau)\,\,\frac{d~}{d\tau}{x}_{_{\mathcal{C}}}\bb{\tau},
\label{zzvquaz62CC}
\end{eqnarray}
and
\begin{eqnarray}
\frac{D~}{D\tau}\mathcal{P}_{(\mathcal{C})}
\bigg{\vert}_{\tau = T} &=& -\oint_{\mathcal{C}}d\ell\, \mathcal{W}\,\Delta\mbox{\boldmath$\mathcal{J}$}\cdot \mathbf{n} \nonumber\\&=& -
\int_{0}^{T}d\tau\, \mathcal{W}(x_{_{\mathcal{C}}}\bb{\tau},\,k_{_{\mathcal{C}}}\bb{\tau};\tau) \,\Delta \mathcal{J}_p(x_{_{\mathcal{C}}}\bb{\tau},\,k_{_{\mathcal{C}}}\bb{\tau};\tau)\,\,\frac{d~}{d\tau}{x}_{_{\mathcal{C}}}\bb{\tau},
\label{zzvquaz64DD}
\end{eqnarray}
for von Neumann entropy and purity, respectively, through which one can quantify the role of quantum fluctuations driven by the contributions from $\Delta\mbox{\boldmath$\mathcal{J}$}\cdot \mathbf{n} \equiv \mbox{\boldmath$\mathcal{J}$}\cdot \mathbf{n}$\footnote{For quantum systems which account for the all order corrections from Eq.~\eqref{zzvDimWB}, quantumness and classicality can be quantified through the above obtained continuity equation framework through the results from Eqs.~\eqref{zzvquaz51EE}, \eqref{zzvquaz62CC}, and \eqref{zzvquaz64DD} \cite{EPL18,JCAP18,NovoPaper}.}.

Of course, as emphasized, the integrals from Eqs.~\eqref{zzvquaz51DD}, \eqref{zzvquaz62CC} and \eqref{zzvquaz64DD} vanish in the classical limit, i.e. for $\mbox{\boldmath$\mathcal{J}$} \sim \mathbf{v}_{\xi(\mathcal{C})}\mathcal{W}$. Therefore the obtained results work as an optimized quantifier of non-classicality for a pletora of Wigner functions.

\section{R\'enyi entropy continuity equation}

\hspace{1 em} The R\'enyi entropy, $R_\beta $, generalizes the Hartley entropy, the Shannon entropy, the collision entropy and the minimal entropy \cite{Jizba}. Once contextualized in the phase-space formulation, it can be written as
\begin{eqnarray}
R_\beta = \frac{1}{1-\beta}\log
\left[
\int^{+\infty}_{-\infty} \hspace{-.15cm}\int^{+\infty}_{-\infty} \hspace{-.15cm} {dx\,dk}\,\mathcal{W}_{(x, k)}^\beta
\right],
\end{eqnarray}
where the index $\beta$ is introduced to denote an estimative of the fractal dimension.
By following the same derivative structure reported in the previous section, one could notice that
\begin{eqnarray}
e^{[(1-\beta)R_\beta]}\frac{D}{D\tau}{R_\beta }\bigg{\vert}_{\tau = T} &=& -\frac{D~}{D\tau}\left(\int_{V}dV\, \mathcal{W}^{\beta}\right)
\nonumber\\
&=& -\int_{V}dV\,\left[\frac{D~}{D\tau} (\mathcal{W}^{\beta}) + \mathcal{W}^{\beta} \mbox{\boldmath $\nabla$}_{\xi}\cdot \mathbf{v}_{\xi(\mathcal{C})}\right]\nonumber\\
&=&- \int_{V}dV\,\left[\frac{\partial~}{\partial \tau} (\mathcal{W}^{\beta}) + \mbox{\boldmath $\nabla$}_{\xi}\cdot(\mathbf{v}_{\xi(\mathcal{C})} \mathcal{W}^{\beta})\right]\nonumber\\
&=&- \int_{V}dV\,\left[-\beta\mathcal{W}^{\beta-1} \mbox{\boldmath $\nabla$}_{\xi}\cdot\mbox{\boldmath$\mathcal{J}$} + \mbox{\boldmath $\nabla$}_{\xi}\cdot(\mathbf{v}_{\xi(\mathcal{C})} \mathcal{W}^{\beta})\right]\nonumber\\
&=&\int_{V}dV\,\left[(\beta-1)\mathcal{W}^{\beta} \,\mbox{\boldmath $\nabla$}_{\xi}\cdot\mathbf{w} 
+ \mbox{\boldmath $\nabla$}_{\xi}\cdot(\mbox{\boldmath$\mathcal{J}$}\mathcal{W}^{\beta-1} - \mathbf{v}_{\xi(\mathcal{C})} \mathcal{W}^{\beta})\right]\nonumber\\
 &=& \int_{V}dV\,(\beta-1)\mathcal{W}^{\beta} \,\mbox{\boldmath $\nabla$}_{\xi}\cdot\mathbf{w} 
+ \oint_{\mathcal{\,\,}}d\ell\, \mathcal{W}^{\beta-1}\left(\mbox{\boldmath$\Delta\mathcal{J}$}\cdot \mathbf{n}\right)\nonumber\\
&\stackrel{V\to\infty}{=}&({\beta-1}) \langle \mathcal{W}^{\beta-1} \mbox{\boldmath $\nabla$}_{\xi}\cdot\mathbf{w}\rangle,
\label{zzvquaz69}
\end{eqnarray}
and that
\begin{eqnarray}
e^{[(1-\beta)R_\beta]}\frac{D}{D\tau}{R_\beta }\bigg{\vert}_{\tau = T}
 &=& \hspace{-.2 cm}-\oint_{\mathcal{C}}d\ell\, \mathcal{W}^{\beta-1}\,\Delta\mbox{\boldmath$\mathcal{J}$}\cdot \mathbf{n} \nonumber\\&=& \hspace{-.2 cm}-
\int_{0}^{T}d\tau\, \mathcal{W}^{\beta-1}(x_{_{\mathcal{C}}}\bb{\tau},\,k_{_{\mathcal{C}}}\bb{\tau};\tau) \,\Delta \mathcal{J}_p(x_{_{\mathcal{C}}}\bb{\tau},\,k_{_{\mathcal{C}}}\bb{\tau};\tau)\,\,\frac{d~}{d\tau}{x}_{_{\mathcal{C}}}\bb{\tau},
\label{zzvquaz64DD}
\end{eqnarray}
which is consistent with the previously obtained continuity equations.

\section{Conclusions}

\hspace{1 em} In this work, it has been shown that the Weyl transform quantum states lead to phase-space continuity equations which allow for evaluating departures from Liouvillian properties of Wigner currents and fluxes.
The results provide the tools for discussing quantumness and classicality in terms of phase-space quantum decoherence, purity and von Neumann entropy fluxes, in the context of (an)harmonic periodic quantum systems.
The informational content analysis was extended in order to provide an equivalent formulation for the phase-space R\'enyi entropy, in terms of an additional continuity equation.
Our results suggest that the Wigner flow framework can be universally applied to distinguish how quantum and classical regimes are quantitatively distinguished from each other \cite{EPL18,JCAP18,NovoPaper}.

\ack{The work of A. E. B. was supported by the Brazilian agencies FAPESP (grant 2018/03960-9) and CNPq (grants 451081/2018-8 and 300831/2016-1). The work of O. B. was partially supported by the COST action MP1405 Quantum Structure of Spacetime (QSPACE).}

\smallskip

\section*{References}
\begin{thebibliography}{9}
\bibitem{Wigner}
Wigner E 1932 {\it Phys. Rev.} {\bf 40} 749
\bibitem{Case}
Case W B 2008 {\it Am. J. Phys.} {\bf 76} 937 
\bibitem{01A}
Bernardini A E and Chinaglia M 2015 {\it Mod. Phys. Lett.} {\bf A30} 1550118
\bibitem{02A}
Steuernagel O, Kakofengitis D and Ritter G 2013 {\it Phys. Rev. Lett.} {\bf 110} 030401 
\bibitem{03A}
Meekhof D M, Monroe C, King B E, Itano W M and Wineland D J 1996 {\it Phys. Rev. Lett.} {\bf 76} 1796
\bibitem{Moyal} 
Moyal J 1949 {\it Proc. Camb. Phil. Soc.} {\bf 45} 99
\bibitem{Ballentine}
Ballentine L E 1998 {\it Quantum Mechanics: a Modern Development} pp. 633 (World Scientific).
\bibitem{EPL18}
Bernardini A E and Bertolami O 2017 {\it EPL} {\bf 120} 20002
\bibitem{Ferraro11}
Ferraro A and Paris M G A 2012 {\it Phys. Rev. Lett.} {\bf 108} 260403
\bibitem{Donoso12}
Donoso A and Martens C C 2001 {\it Phys. Rev. Lett.} {\bf 87} 223202
\bibitem{04}
Domcke W, H\"{a}nggi P and Tannor D 1997 {\it Chem. Phys.} {\bf 217} 117
\bibitem{JCAP18}
Bernardini A E, Leal P and Bertolami O 2018 {\it JCAP} {\bf 02} 025
\bibitem{Jizba}
Jizba P and Arimitsu T 2004 {\it Annals of Physics} {\bf 312} 17
\bibitem{NovoPaper}
Bernardini A E 2018 {\it Phys. Rev.} {\bf A98} 052128
\end{thebibliography}
\end{document}